\documentclass{article}
\usepackage{here}
\usepackage{epsfig}
\newcommand{\IGN}[1]{}
\begin{document}
\pagestyle{myheadings}
\title{\vspace*{-20mm}\mbox{\mbox{} \hspace*{24mm}\normalsize Published in: \underline{Lecture Notes in Computer Science {\bf 4131}, 208-215 (2006).}}\\[4mm]
Investigation of topographical  stability of the concave and convex
Self-Organizing Map variant}
\author{Fabien Molle$^{1,2}$ and Jens Christian Claussen$^2$\footnote{Corresponding author.} 
\\[6mm]\small 
$^1$Theoretical Physics,
Chalmers Tekniska H\"ogskola, 
G\"oteborg
\\
\small
$^2$Institut f\"ur Theoretische Physik und Astrophysik
\\\small
Leibnizstr. 15, 24098 Christian-Albrechts-Universit\"at zu Kiel, Germany
\\\small
\small \texttt{http://www.theo-physik.uni-kiel.de/\~{ }claussen/} \\[-3mm] } \date{}
\maketitle        
\setcounter{page}{208}
\pagestyle{myheadings} 
\markboth{F.\ Molle and J.\ C.\ Claussen \hfill  Investigation of topographical stability ~~~ }  {F.\ Molle and J.\ C.\ Claussen \hfill  Investigation of topographical stability ~~~}

%


%
%
%
%
%
%
\begin{abstract}
We investigate, by a systematic numerical study, the parameter dependence 
of the stability of the Kohonen Self-Organizing Map
and the Zheng and Greenleaf concave and convex learning
with respect to different input distributions, 
input and output dimensions.
\\[1ex]
{\small\sl
Topical groups: Advances in Neural Network Learning Methods,
Neural and hybrid architectures and learning algorithms,
Self-organization.\\[-3.5ex]
}
\end{abstract}
\noindent
Neural vector quantizers have become a widespreadly 
used tool to explore high-dimensional data sets
by self-organized learning schemes.
Compared to the vast literature on variants and applications
that appeared the last two decades, the theoretical description
proceeded more slowly.
Even for the coining Self-Organizing Map (SOM) 
\cite{kohonen82},
still open questions remain, as a proper 
description of the dynamics for the case of dimension reduction
and varying data dimensionality, or the
question for what parameters stability of the algorithm
can be guaranteed.
This paper is devoted to the latter question.
The stability criteria are especially interesting for
modifications and variants, as the concave and convex learning
\cite{zheng96a},
whose magnification behaviour has been discussed recently 
\cite{villmannclaussen06}.
Especially for the variants, analytical progress
becomes quite 
difficult,
and in any case one will expect that the
stability will depend 
on the input distribution
to some ---apart from special cases--- unknown extent.
As the invariant density 
in general
is analytically unaccessible
for input dimensions larger than one
(see \cite{kohonen99} for recent tractable cases),
we expect a general theory not to be available immediately,
and instead proceed with a systematic numerical exploration.
\\[-5ex]

\paragraph*{\sl The Kohonen SOM, and the nonlinear variant of Zheng and Greenleaf. --}
The class of algorithms investigated here is defined 
by the learning rule,
that for each stimulus  ${\bf v} \in V$
 each weight vector ${\bf w}_{{\bf r}}$ is updated according to
\\[-4ex]
        \begin{eqnarray}
        {\bf w}_{{\bf r}}^{{\sf new}} =
         {\bf w}_{{\bf r}}^{{\sf old}}
         + \varepsilon \cdot g_{{\bf r} {\bf s}}
         \cdot ({\bf v}-{\bf w}_{{\bf r}}^{{\sf old}})^{K}
        \end{eqnarray}
\\[-4ex]
($g_{{\bf r}{\bf s}}$ being a gaussian function (width $\sigma$) of euclidian
distance $|{\bf r}-{\bf s}|$ in the neural layer,
thus describing the neural topology).
Herein
\\[-4ex]
\begin{eqnarray}
\mbox{$|{\bf w}_{{\bf s}} -{\bf v}|$} =
\mbox{$\min_{{\bf r}\in R}
|{\bf w}_{{\bf r}} -{\bf v}|$} \label{eq:voronoi}
\end{eqnarray}
\mbox{}
\\[-4ex]
determines for each stimulus ${\bf v}$ 
the {\sl best-matching unit} or {\sl winner} neuron.
\clearpage

The case $K=1$ is the original SOM \cite{kohonen82},
corresponding to a linear or Hebbian learning rule.
The generalization to $K$ or $1/K$ taking integer values 
has been proposed by 
Zheng and Greenleaf
\cite{zheng96a},
but arbitrary nonzero real values of $K$ can be used
\cite{villmannclaussen06},
and the choice of $K\to{}0$ has been shown 
(for the onedimensional
case) to
have an invariant density with the information-theoretically 
optimal value of the magnification exponent one
\cite{villmannclaussen06}, 
i.e., the neural density is proportional to the input density
and hence can be used as a density estimator.
\\[-5ex]

\paragraph*{\sl Convergence and stability. --}
It is well known that for the learning rate $\varepsilon$,
one has to fulfill the Robbins-Munro conditions
(see, e.g.\ \cite{kohonen99})
to ensure convergence, with all other parameters fixed.
However, practically it is necessary to use a large
neighborhood width at the beginning, to have the
network of weight vectors ordered in input space,
and decrease this width in the course of time
downto a small value that ensures topology preservation
during further on-line learning.
Thus the situation becomes more involved when additionally
also $\sigma$ is made time-dependent.
Here we consider the strategy where the stability border in
the $(\varepsilon,\sigma)$ plane always is approached 
from small $\varepsilon$ with $\sigma$ fixed during this final phase.
An ordered state has to be
generated by preceding learning phases.
\\[-5ex]

\paragraph*{\sl Measures for Topographical Stability. --}
To quantify the ordered state and the topology preservation,
a variety of measures is used, 
e.g.\ the topographic product \cite{villmann97},
the Zrehen measure \cite{zrehen1993},
and the average quadratic reconstruction error.
To detect instable behaviour, all measures should be
suitable and give similar results.
For an unstable and disordered map, also the total sum over all
(squared) distances between adjacent weight vectors
will increase significantly; so a threshholded increase 
will indicate instability as well. 
This indicator is used below; however, for the
case of a large neighborhood (of network size), 
the weight vectors shrink to a small volume, 
thus influencing the results; however, this 
applies to a neighborhood widths larger than that
commonly used for the pre-ordering.

\begin{figure}[H]
\hspace*{0.1\textwidth} 
\hspace*{0.1\textwidth}
\epsfig{file=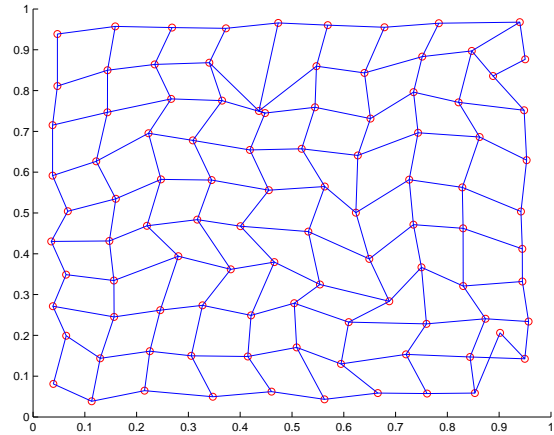,width=0.58\textwidth}
\caption{
Situation where
one defect is detected by the crossproduct measure
(\ref{crossproduct}).
\label{figOVER}}
\end{figure}

In addition we use here an even more simple approach than the
Zrehen measure (which counts the number of
neurons that lie within a circle between 
each pair of neurons that are adjacent in the neural layer).
For a mapping from $d$ to $d$ dimensions, 
we consider the determinant of the
$i$ vectors spanned by 
$\vec{w}_{\vec{r}+\vec{\rm e}_i}-\vec{w}_{\vec{r}}$,
with
$\vec{\rm e}_i$ being the $i^{\rm th}$
unit vector.
The sign of this determinant, where $1\leq i \leq d$,
thus gives the orientation of the set of $d$ vectors.
Note that the 1-dimensional case just reads
${\rm sgn}(w_{r+1}-w_{r})$, which has been widely considered
to detect the ordered state in the 1 to 1 dimensional case.
Hence, we can define 
\\[-4ex]
\begin{eqnarray}
\label{crossproduct}
\chi (\{ \vec{w}_{\vec{r}} \}) :=
~~~~~~~~~~~~~~~~~~~~~~~~~~~~~~~~~~~~~~~~~~
~~~~~~~~~~~~~~~~~~~~~~~~~~~~~~~~~~~~~~~~~~
\\
 1 - \frac{1}{N} \left|\sum_{\vec{r}}
{\rm sgn} ( \det(
(\vec{w}_{\vec{r}+\vec{\rm e}_1}-\vec{w}_{\vec{r}}),
\ldots
(\vec{w}_{\vec{r}+\vec{\rm e}_i}-\vec{w}_{\vec{r}}),
\ldots
(\vec{w}_{\vec{r}+\vec{\rm e}_d}-\vec{w}_{\vec{r}})
))\right|.
\nonumber
\end{eqnarray}
This evaluates the number of neurons 
$N_+$ (resp.\ $N_-$),
where this sign is positive (resp.\ negative),
hence the relative fraction
of minority signs is given by
$(1- |N_+-N_-|/N)$.
A typical single defect is shown in Fig.~\ref{figOVER}.
Due to its simplicity, this measure 
$\chi$ will be used in
the remainder.
\\[-5ex]

\paragraph*{\sl Modification of learning rules and data representation. --}
A classical result \cite{ritter86,ritter88,ritter92}
states that the neural density for 1-D SOM (in the continuum limes)
approaches not the input density itself, but a power of it,
with exponent $2/3$, the so-called magnification exponent.
As pointed out by Linsker \cite{Linsker89a},
the case of an exponent 1 would correspond to the case of 
maximal mutual information between input and output space.
Different modifications of the 
winner mechanism or the learning rule, by additive or multiplicative
terms, have been suggested and 
influence the magnification exponent
\cite{heskes99a,erwin92b,claussen05,claussenvillmann05}.
Here we investigate the case of 
concave and convex learning
\cite{zheng96a,villmannclaussen06},
which defines a nonlinear generalization of the SOM.
\\[-4ex]

\begin{figure}[hptb]
\raisebox{-15mm}{\epsfig{file=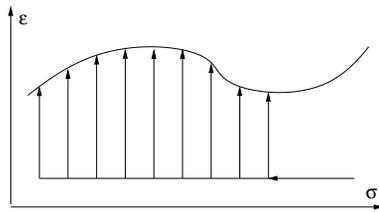,width=0.4\textwidth}}
\raisebox{0mm}{\begin{minipage}{0.58\textwidth}
\caption{Schematic diagram of the parameter path
in $(\varepsilon,\sigma)$ space.
Starting with high values, $\sigma$ is 
slowly decreased to the desired value, while the
learning rate still is kept safely low.
From there, at constant $\sigma$ the learning rate is
increased until instability is observed; giving an 
upper border to the stability area.
-- 
The same scheme is applied for the concave and convex
learning, where the nonlinearity exponent is
considered as a fixed parameter.
\label{figPATHscheme}}
\end{minipage}}
\\
\end{figure}

\begin{figure}[H]
\centerline{
\epsfig{file=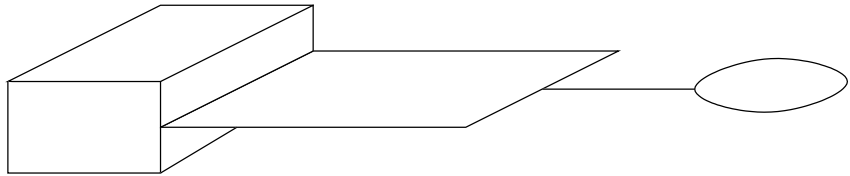,width=0.9\textwidth}
}
\caption{
Schematic view of the classical box-plane-stick-loop dataset.
Its motivation is to combine locally different input data dimensions
within one data set.
\label{figBPSL}}
\end{figure}

\begin{figure}[H]
\hspace*{0.1\textwidth} 
\epsfig{file=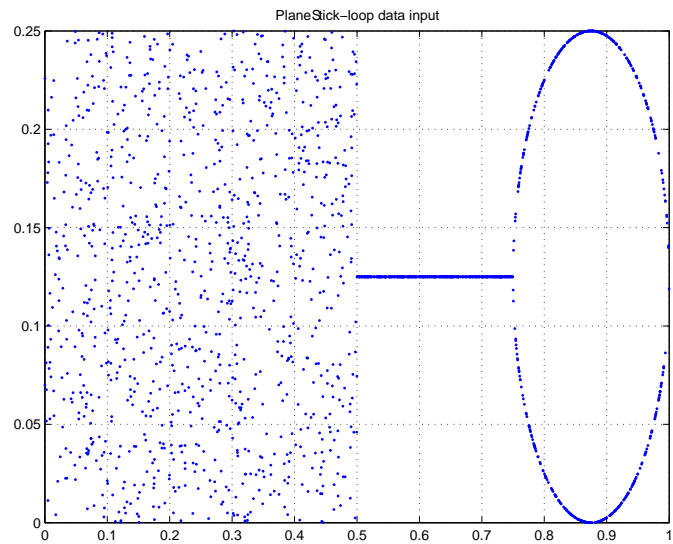,width=0.78\textwidth} 
\caption{Part of the input data
for the
2D 
plane-stick-loop data set 
(Fig.\ \ref{figBPSL}).
\label{figBPSLdata}}
\end{figure}

\begin{figure}[H] 
\begin{center}
\epsfig{file=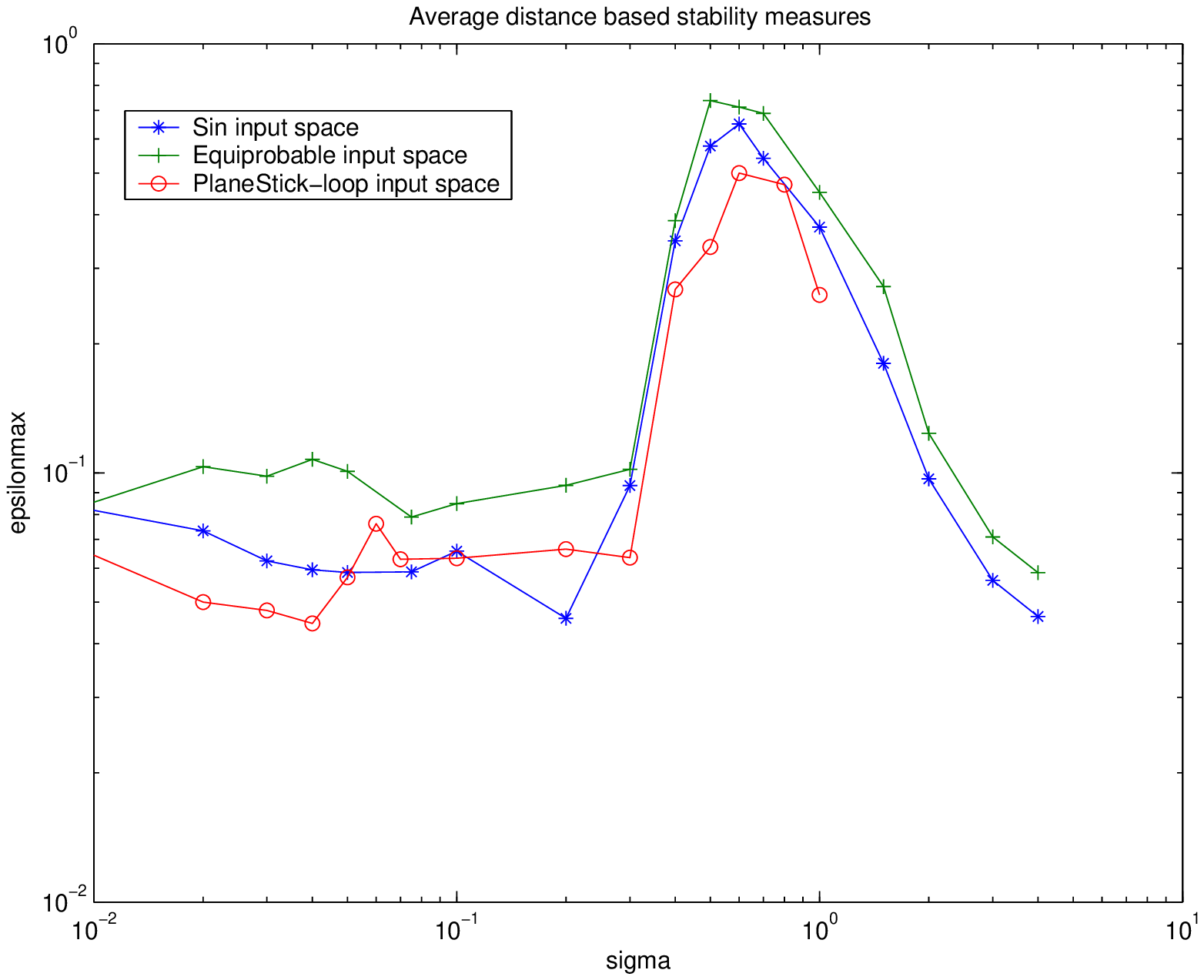,height=0.345\textheight}
\\
\epsfig{file=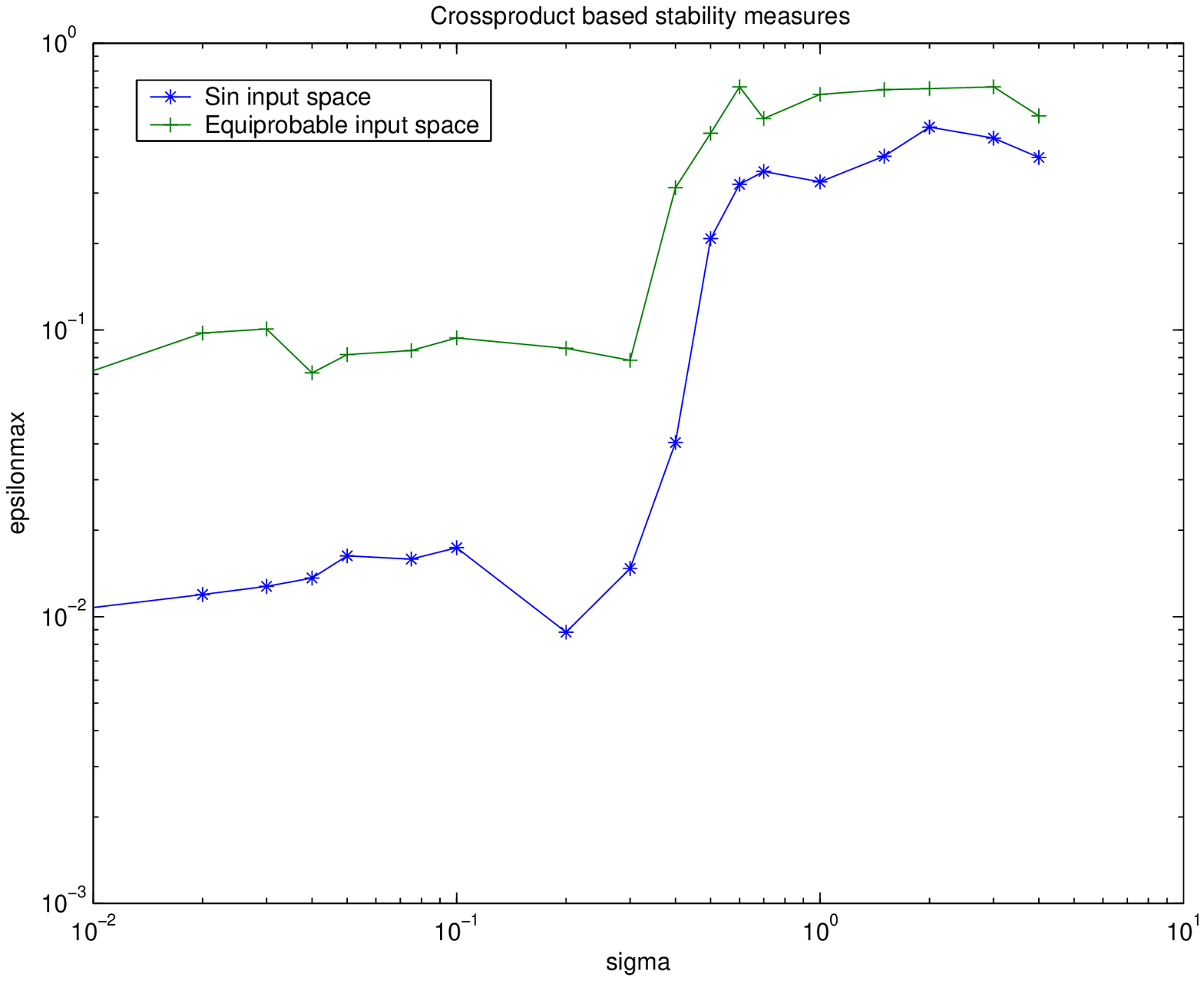,height=0.33\textheight}
\end{center}
\vspace*{-2ex}
\caption{
Critical $\varepsilon_{\rm max}(\sigma)$ where
(coming from small $\varepsilon$ values,
see Fig.\ \ref{figPATHscheme}) the
SOM learning loses stability.
Here a 2D array of 10$\times$10 neurons was used
with decay $\exp(-t/k)$ exponentially in time $t$,
with $k$ between 30000 and 60000
depending on $\varepsilon_0$ (for $\sigma$ between
0.1 and 0.001, $k=300000$).
Top: Unstable $\varepsilon$
detected from growth of the averaged distance of neurons;
here a threshold of $15\%$ was chosen.
For large $\sigma$, this measure becomes less reliable 
due to shrinking of the network, i.e.\
 $\forall_{\vec{r}} \vec{w}_{\vec{r}}\to \langle\vec{v}\rangle$.
Bottom: Unstable $\varepsilon$ 
detected from the crossproduct measure,
eq.\ (\ref{crossproduct}),
with threshold of 1 defect per 100 iterations.
The $\varepsilon$ value depends on the data 
distribution, but the qualitative behaviour remains similar.
In all cases, below a certain 
$\sigma_{\rm crit}$ of about 0.3, 
$\varepsilon$ has to be decreased significantly.
This independently reproduces 
\cite{villmannthesis}, 
here we investigate also different $\varepsilon$ values.
\label{figSOM}}
\end{figure}

\paragraph*{\sl Topographical Stability for the Self-Organizing Map. --}
Before investigating the case of concave and convex learning,
the stability measures should be tested for the
well-established SOM algorithm. 
Using the parameter path of Fig.~\ref{figPATHscheme},
we first analyze the
2D $\to$ 2D case, 
for three input distributions: the
homogeneous input density (equidistribution), an
inhomogeneous input distribution
$\sim\sin(\pi x_i)$
\cite{claussenvillmann05},
and a varying-dimension dataset 
(Figs.\ \ref{figBPSL}, \ref{figBPSLdata}).
The results are shown in Fig.~\ref{figSOM}.
\\[-4ex]

\begin{figure}[H]
\begin{center}
\epsfig{file=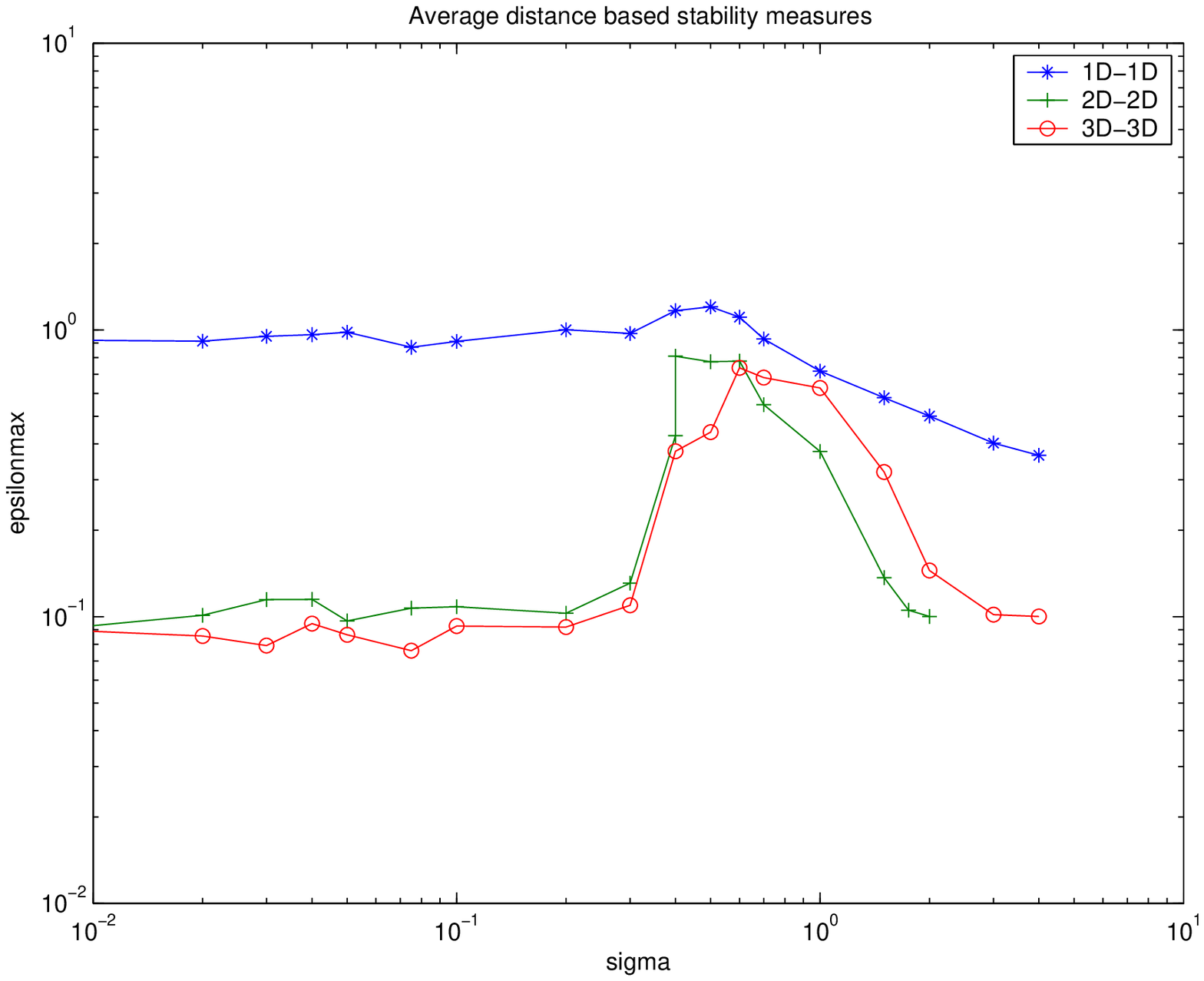,width=0.7\textwidth}
\epsfig{file=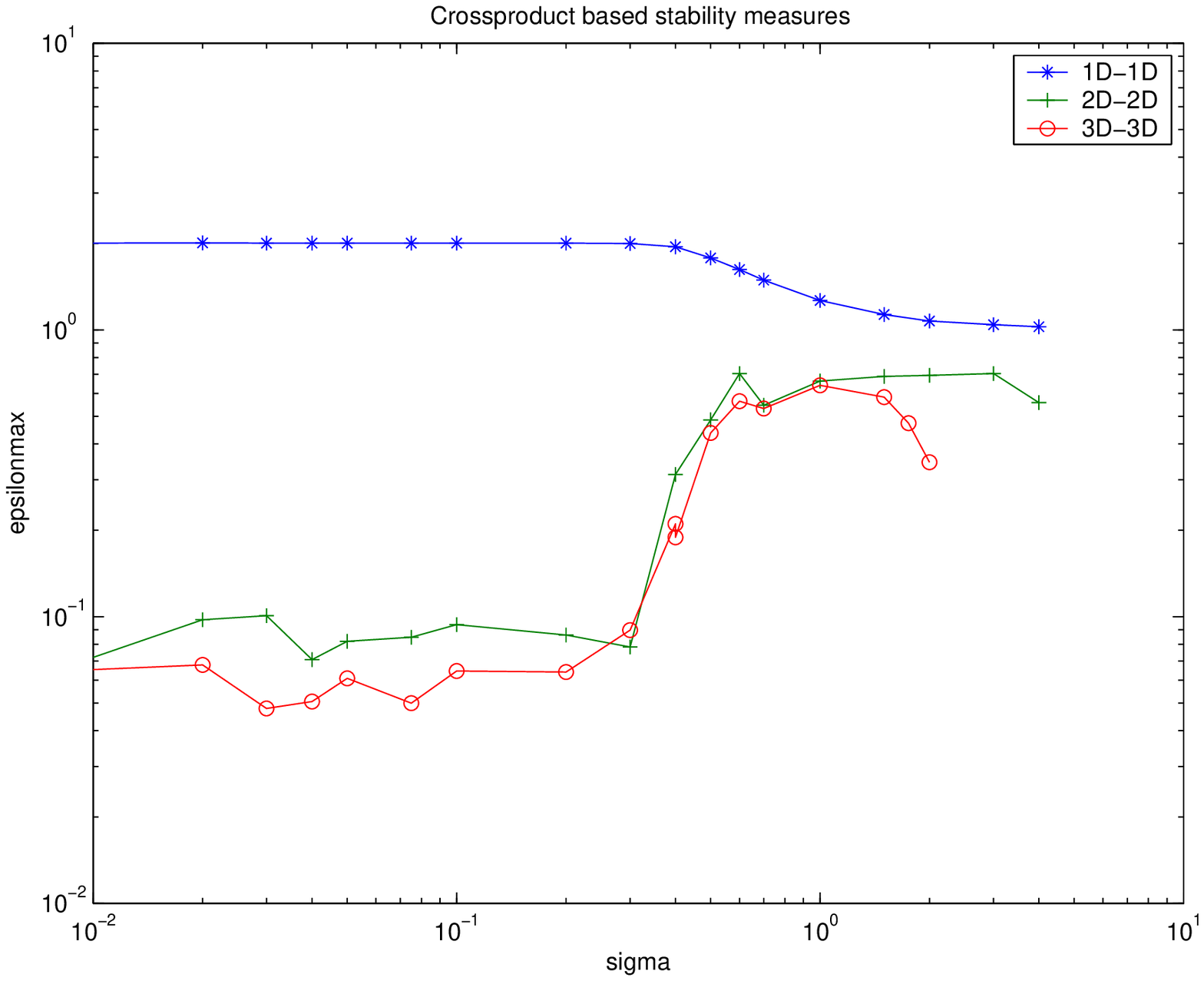,width=0.7\textwidth}
\end{center}
\caption{
Stability border dependence on input dimension
(1D, 2D, 3D).
The known 1D case is included for comparison.
Top: Using the average length criterion
(for $\sigma>1$, the result can be misleading
due to total shrinking of the network, see text).
Bottom: Using the crossproduct detection for defects,
similar results are obtained; for large $\sigma$
instabilities are detected earlier.
\label{fig123D}}
\end{figure}

\vspace*{-2ex}
\paragraph*{\sl Different input dimensions
and varying intrinsic dimension. --}
As the input dimensionality is of pronounced influence on
the maximal stable learning rate
(Fig.\ \ref{fig123D}), we also investigate 
an artificial dataset combining different dimensions:
the box-plane-stick-loop dataset 
\cite{martinetz93d}
(Fig. \ref{figBPSL}), or its 
2D counterpart, the plane-stick-loop
(Fig. \ref{figBPSLdata}).
Here the crossproduct detection will become problematic 
where the input space is intrinsically 1D (stick and loop),
thus the average distance criterion is used, and
we restrict to the case $\sigma\leq 1$.

\clearpage

\clearpage

\begin{figure}[H]
\begin{center}
\epsfig{file=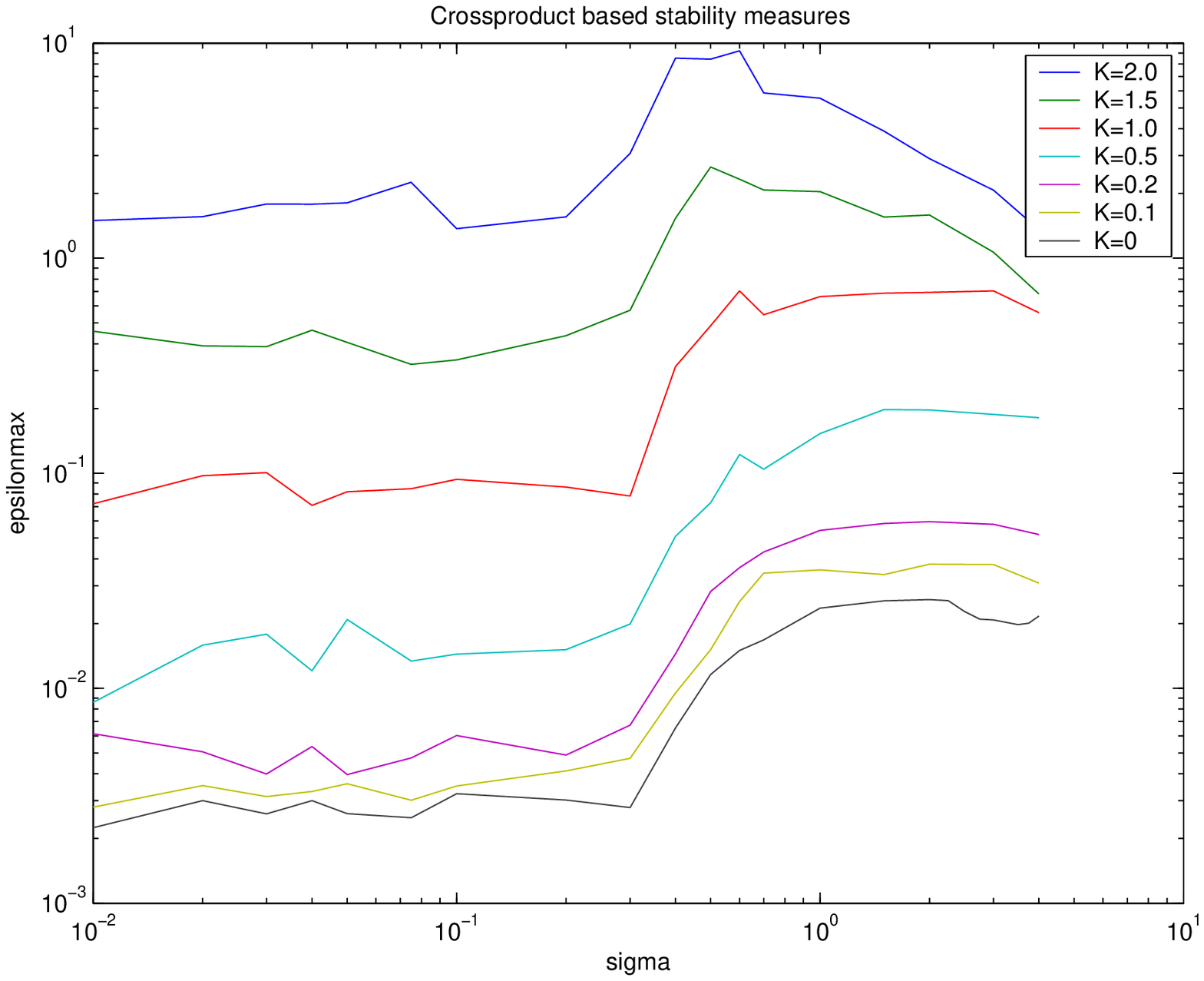,width=0.7\textwidth}
\\
\epsfig{file=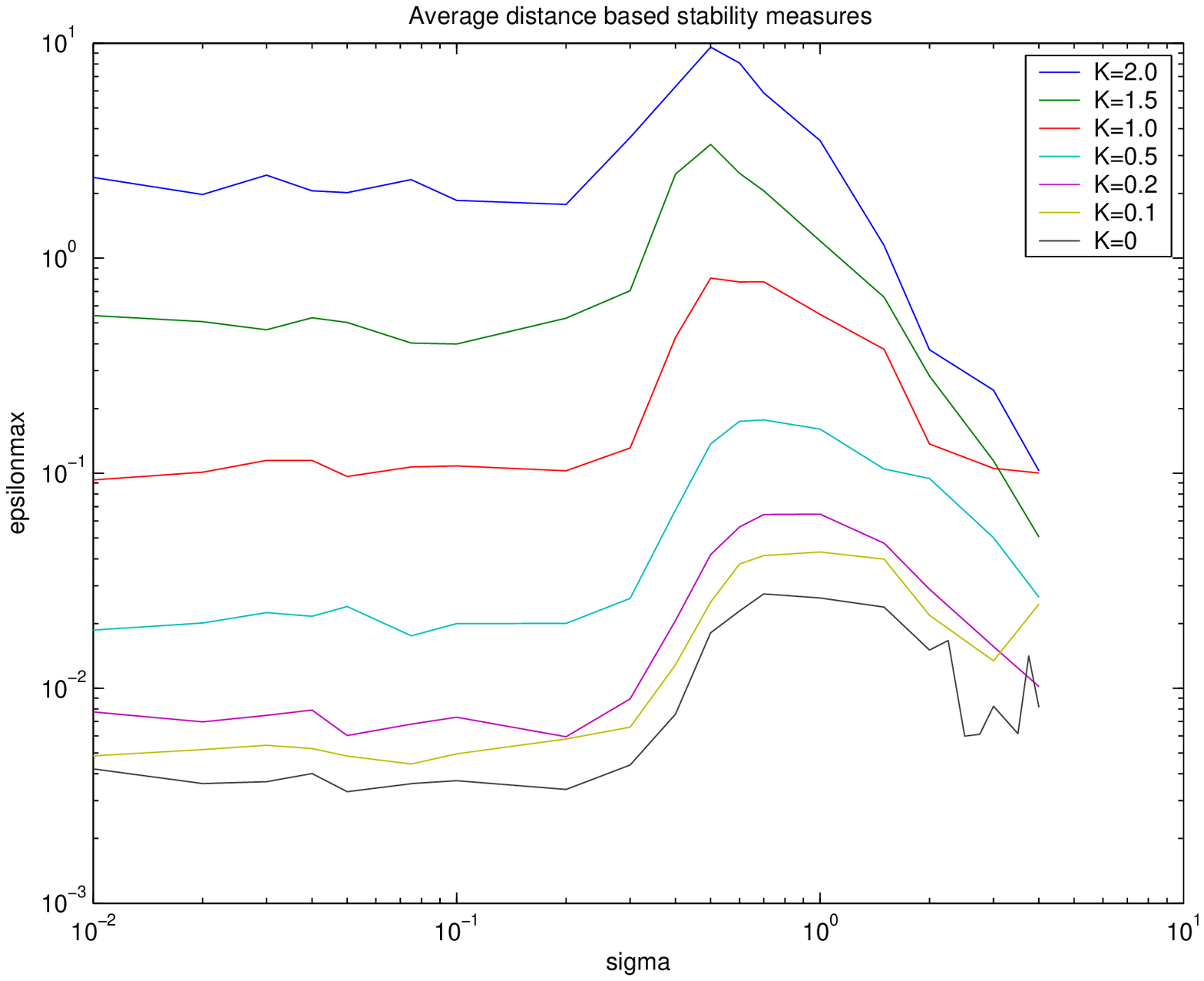,width=0.7\textwidth}
\end{center}
\caption{
Critical $\varepsilon_{\rm max}(\sigma)$ 
for different values of the nonlinearity parameter
from $K=2.0$ (top) to $K=0$ (bottom).
$K=1$ corresponds to the SOM case.
\label{figCONV}}
\end{figure}

\clearpage
\paragraph*{\sl Concave and convex learning: 
Stability of nonlinear learning. --}
The simulation results
are given in Fig.~\ref{figCONV}:
Clearly, a strong influence of the nonlinearity
parameter $K$ is observed. 
Especially one has to take care when decreasing 
$\sigma$, because for too large $\epsilon$
the network becomes instable. 
For $K<1$, much smaller values of $\epsilon$
are possible, thus considerably longer learning
phases have to be taken into account compared to
original SOM.
For $K>1$ the stability range becomes larger.

\clearpage
\paragraph*{\sl Discussion. --}
We have defined a standardized testbed for 
the stability analysis of SOM vector quantizers 
with serial pattern presentation,
and compared the SOM with the recently
introduced variants of concave and convex learning.
The stability regions 
for different input distribution and dimension are
 of the same shape, thus
qualitatively similar, but not coinciding exactly.
The neighborhood 
width, but unfortunately also the input 
distribution affect the maximal stable learning rate.
For the concave and convex learning, 
the exponent steering the nonlinear learning
also crucially influences the learning rate.
In all cases, a plateau for $\sigma \ll 1$ is found
where the learning rate  must be quite low
compared to the intermediate range 
$0.3 \leq \sigma \leq 1$.
As a too safe choice of the learning rate 
simply increases computational cost,
an accurate knowledge of the stability
range of neural vector quantizers 
is of direct relevance in many applications.
\\[-4ex]

\end{document}